\documentclass[10pt]{article}
\usepackage{amsfonts}
\usepackage{amsmath}
\usepackage{amssymb}
\usepackage{graphicx}

\def \be {\begin{equation}}
\def \ee {\end{equation}}
\def \bea {\begin{eqnarray}}
\def \eea {\end{eqnarray}}

\def \rr {\raise.35ex\hbox{\small $\prime$}\kern-.17em{\mbox{\large $\imath$}}}
\def \del {\partial}
\def \dels {\partial\kern-.5em / \kern.5em}
\def \As {{A\kern-.5em / \kern.5em}}
\def \Ds {D\kern-.7em / \kern.5em}

\def \a {\alpha}

\def \ep {\epsilon}
\def \m {\mu}
\def \n {\nu}

\def \lam {\lambda}

\begin{document}
\begin{titlepage}
%\catcode`\@=11
%\catcode`\@=12
%\twocolumn[\hsize\textwidth\columnwidth\hsize\csname%
%@twocolumnfalse\endcsname

%\draft
\begin{center}
\hfill hep-th/0604026\\
\vskip .5in

\textbf{\Large
Linear relations among
4-point functions \\
\vskip .1in
in the high energy limit
of string theory}

\vskip .5in
{\large Pei-Ming Ho, Xue-Yan Lin}

\vskip 15pt

{\small Department of Physics,
National Taiwan University,
Taipei, Taiwan, R.O.C.}

\vskip .2in
\sffamily{
pmho@phys.ntu.edu.tw \\
xueyan.lin@msa.hinet.net}

\vspace{60pt}
%\maketitle
\end{center}
\begin{abstract}
The decoupling of zero-norm states
leads to linear relations among
4-point functions
in the high energy limit
of string theory.
Recently it was shown that
the linear relations uniquely determine
ratios among 4-point functions
at the leading order.
The purpose of this paper is
to extend the validity of the same approach
to the next-to-leading order and higher orders.
\end{abstract}
%\pacs{PACS numbers: 11.25.-w, 11.25.Mj, 11.25.Sq}%]
\end{titlepage}
%\begin{narrowtext}
\setcounter{footnote}{0}

\section{Motivation}

Many have posed the question whether
there is a fundamental principle underlying string theory,
in a way analogous to how the equivalence principle
led Einstein to general relativity.
For example, for the bosonic open string theory,
Witten's cubic string field theory is formally
a Chern-Simons theory
with infinite gauge symmetry,
but the gauge symmetry of the infinitely many
massive higher spin gauge fields is not manifest
in the flat background.
It is tempting to imagine that a Higgs mechanism
is responsible for the masses of the gauge fields,
and the symmetry will be restored when
we consider physics at sufficiently high energies.
Gross and Mende \cite{GrossMende}, Gross \cite{Gross}
and Gross and Manes \cite{GrossManes}
explored this possibility for bosonic open and closed strings.
They used the saddle point approximation to
evaluate the integral over moduli space
for 4-point functions.
They found that, to the 0th order approximation
in the high energy limit,
the saddle point is independent
of the particles participating the scattering,
and claimed that this is a signal of the hidden symmetry.
Unfortunately,
it was shown later \cite{ChanLee1,ChanLee2,CHL}
that some of their results are incorrect.
On the other hand,
it is not clear how the existence of
a universal 0th order saddle point
is related to the existence of symmetry.

Following a series of earlier works
\cite{ChanLee1}-\cite{CHLTY4},
our approach is to explore the implication of
the decoupling of %spurious states
zero-norm states
from other physical states.
%Spurious states
Zero-norm states correspond to gauge symmetries.
One of the most salient features of string theory
is that it has infinitely many higher spin 
gauge fields with infinitely many gauge symmetries.
There is a delicate conspiracy
among all the fields and symmetries so that,
in spite of many no-go theorems and folklore
about interacting higher spin gauge fields,
string theory is a consistent interacting theory
of massive higher spin gauge fields in flat spacetime.
In view of this mystery
it is worthwhile to study %spurious states 
zero-norm states in detail.

In \cite{ChanLee1}-\cite{CHL}
and \cite{CHLTY1}-\cite{CHLTY4},
the decoupling of
%spurious states
zero-norm states is used to
derive linear relations among 4-point functions
at the leading order in the high energy limit.
In the old covariant first quantized formulation
of string theory,
%spurious states
zero-norm states represent
on-shell gauge transformations.
Normally, on-shell gauge transformations
only transform a state to itself in a different gauge,
and never transform a state to another different state.
The crucial step which made it possible to relate
physically inequivalent states is that,
in the high energy limit,
we ignore subleading terms in the gauge transformation,
and the %spurious states
zero-norm states are no longer exactly
orthogonal to all physical states.
It turns out that the many gauge transformations
intertwine and overlap with each other so much that
the assumption of a smooth, consistent high energy limit
of string theory uniquely fixes ratios among
4-point functions to the leading order.
Remarkably, via simple algebraic manipulations,
numerical ratios among 4-point functions
were obtained explicitly for all mass levels
\cite{CHLTY1,CHLTY2,CHLTY3}.
These ratios involve all 4-point functions
at the leading order in the high energy limit.

The purpose of this paper is to extend
our understanding to the next-to-leading order.
The first question is whether amplitudes at
the next-to-leading order are also unique up to
an overall constant at all mass levels.
Our answer is Yes.
We also claim that there are linear relations
at higher orders (See Sec.\ref{comment}),
although they are not sufficient to fix
all amplitudes of that order to be proportional
to each other.

There are other approaches in the literature
which are also based on studies of
the algebraic structure
of the string worldsheet theory,
such as \cite{Moore} and \cite{West}.
Our approach distinguishes itself
by giving the simplest and most explicit relations
among correlation functions.
People also tried to define
tensionless strings \cite{WS}
to describe strings in the high energy limit,
as well as to construct various
higher spin gauge theories \cite{HS}
to mimic string theory.
These approaches illuminate different aspects
of the problem and suggest answers in different directions.
Hopefully we will be able to make connections
with these other approaches to have
a better understanding of the hidden symmetry.

\section{Preliminaries and conventions}

The focus of this paper is on
4-point functions at the next-to-leading order.
3-point functions are trivial in the sense
that we can not take a high energy limit
without going off-shell,
and there is no parameter
other than the center of mass energy
so that all amplitudes of the same order
are trivially identical up to a constant factor.
On the other hand,
5-point functions depend on
too many parameters and
so the linear relations are insufficient
to determine their ratios uniquely.

For the scattering process of 2 incoming
and 2 outgoing particles,
the 4 momenta define a 2+1 dimensional scattering plane.
Due to the Poincar\'{e} symmetry,
the scattering amplitude only depends on 2 parameters.
We choose them to be $E$ and $\phi$.
$E$ is the center of mass energy ($s = 4E^2$),
and $\phi$ is the scattering angle
between the momenta of the 1st and 3rd particles
in the center of mass frame.

The scattering amplitude is given by
a 4-point correlation function.
The high energy limit we consider is
\begin{equation}
E\rightarrow \infty, \qquad
\phi = \mbox{fixed}.
\end{equation}
That is, we compute 4-point functions
and keep only the first nonvanishing terms
in the $1/E$ expansion.

For a particle of mass $M$ and momentum
\begin{equation}
P = (E, k, 0, \cdots, 0),
\end{equation}
we will use the following basis of polarizations
\begin{eqnarray}
e^P &=& (E/M, k/M, 0, \cdots, 0), \\
e^L &=& (k/M, E/M, 0, \cdots, 0), \\
e^T &=& (0, 0, 1, \cdots, 0),
\end{eqnarray}
and call them the momentum,
longitudinal and transverse polarizations,
respectively.
These constitute the basis of vectors
on the scattering plane.
We will use $e^I$ to denote
unit vectors perpendicular
to the scattering plane.

When we compare 4-point functions,
we fix 3 of the 4 vertices (say $V_2, V_3, V_4$),
and let only one vertex ($V_1$) be different.
Each polarization vector ($e^T, e^L, e^I$)
of the vertex operator
\begin{equation}
V_1(k_1) = \left[(\del X^L)^m (\del X^T)^n
(\del X^I)^q \cdots\right]
e^{ik_1\cdot X},
\end{equation}
which corresponds to the state
\begin{equation}
(\a^L_{-1})^m (\a^T_{-1})^n (\a^I_{-1})^q
\cdots |0;k\rangle,
\end{equation}
will have to be contracted with another vector
to form a Lorentz invariant in the expression
of the scattering amplitude.
In the high energy limit,
the amplitude is dominated by
contractions with momenta $k_i$.
The polarizations $e^I$ perpendicular
to the scattering plane
is kinematically suppressed as
their contractions with $k_i$'s vanish.
The time-like polarization $e^P$
can also be avoided as a choice of gauge fixing.
Therefore, we will focus our attention
on only two polarizations $e^L$ and $e^T$.

At the leading order in the $1/E$ expansion
of the 4-point function,
in addition to the common factor resembling
the 4-tachyon amplitude,
each factor of $\del^n X^{\mu}$ in the vertex operator
contributes a certain power of $E$
to the scattering amplitude.
We have \cite{CHLTY1}
\begin{eqnarray}
(\a^L_{-1})^{2m} &\sim &E^{2m}, \\
(\a^L_{-1})^{2m+1} &\sim &E^{2m}, \\
\a^L_{-n}&\sim& E^2, \qquad n \geq 2, \\
\a^T_{-n}&\sim &E^1, \qquad n \geq 1.
\end{eqnarray}
In general, when we compare 4-point functions
for different $V_1$'s at the same mass level,
the highest spin state
$(\a^T_{-1})^n|0;k\rangle$
is always at the leading order.
There are more and more other states
at the leading order when we go to higher
and higher mass levels.
They are all of the form \cite{CHLTY1}
\begin{equation}
(\a^L_{-1})^{2m}(\a^T_{-1})^n(\a^L_{-2})^q
|0;k\rangle.
\end{equation}

Ratios between two 4-point functions
at the leading order are uniquely fixed
by the requirement that %spurious states
zero-norm states
be decoupled from all physical states,
assuming that string theory
has a smooth high energy limit.
(The derivation of the ratios is simplied by
considering the decoupling of spurious states.)
The master equation which gives the ratio
between any two leading order amplitude
(for $V_1$'s at the same mass level) is
\cite{CHLTY1}-\cite{CHLTY3}
\begin{equation}
\lim_{E\rightarrow \infty}
\frac{{\cal T}\left(|V_1\rangle=
(\a^T_{-1})^{n-2m-2q}(\a^L_{-1})^{2m}
(\a^L_{-2})^q|0,k_1\rangle\right)}
{{\cal T}\left(|V_1\rangle=
(\a^T_{-1})^n|0;k_1\rangle\right)}
= \left(\frac{-1}{m_1}\right)^{2m+q}
\left(\frac{1}{2}\right)^{m+q}(2m-1)!!.
\label{TheRatio}
\end{equation}
The same result can also be derived from
Virasoro constraints.

The linear relations found before
and those that will be derived here
apply to different choices of $V_1$
at the same mass level,
and are independent of the choices of $V_2, V_3, V_4$.
When we say that an amplitude is at the leading order,
we mean that it is among the most dominant amplitudes
in the high energy limit
for all possible choices of $V_1$
at the same mass level
(without changing $V_2, V_3, V_4$).
Eq.(\ref{TheRatio}) says that the $\phi$-dependence of
the leading order terms in the $1/E$ expansion of
all leading order amplitudes are the same
(for a given mass level),
and the numerical ratios can be uniquely determined.
%by requiring the decoupling of spurious states.

In this paper we focus on the bosonic open string theory.
Our result can be immediately applied to bosonic closed strings,
whose amplitudes factorize into open string amplitudes. 
It should also be possible to extend our results to superstrings.
Definitions of the kinematic variables
of 4-point functions are given in the appendix,
and we will use the convention that $\alpha' = 1/2$.

\section{First massive level ($M^2 = 2$)}

In this section we take the first massive level
as an example to review earlier results
\cite{ChanLee1}-\cite{CHL}.
Readers familiar with these results
should skip to the next section.

Physical states in the first massive level are
\begin {equation}
\label{mass2state}
\left(\ep_{\m\n} \a_{-1}^\m \a_{-1}^\n +
\ep_{\mu}\alpha_{-2}^{\mu}\right) |0;k\rangle,
\end {equation}
where the parameters $\epsilon_{\mu\nu}$,
$\epsilon_{\mu}$ and $k_{\mu}$ satisfy
\begin{equation}
k^2=-M^2=-2,
\qquad
2\ep_{\m\n}k^\n+2\ep_\m=0,
\qquad
\ep^\m{}_\m+2\ep_\m k^\m=0.
\end{equation}
There are two sets of spurious states
\begin{eqnarray}
&L_{-1} \a_{-1}^\m |0;k\rangle
=(p_\n\a^\m_{-1}\a^\n_{-1}+\a^\m_{-2})|0;k\rangle, \\
&L_{-2}|0;k\rangle
=\left(\frac{1}{2} \eta_{\m\n}a^\m_{-1}\a^\n_{-1}
+p_\m\a^\m_{-2}\right)|0;k\rangle.
\end{eqnarray}

Some states are both spurious and physical,
they have zero norm and are
called zero-norm states.
There are two sets of zero-norm states
at this mass level
\begin{eqnarray}
(\Lambda_{(\m}p_{\n)}a^\m_{-1}\a^\n_{-1}
+\Lambda_\m \a^\m_{-2})|0;k\rangle, \qquad
\Lambda\cdot p=0, \label{ZNS11} \\
\lambda\left[(\frac{1}{2}\eta_{\m\n}
+\frac{3}{2}p_\m p_\n)a^\m_{-1}\a^\n_{-1}
+\frac{5}{2}p_\m \a^\m_{-2}\right]|0;k\rangle.
\label{ZNS12}
\end{eqnarray}

The first step of our approach is to
demand zero-norm states to be decoupled from
physical states
\begin{equation}
{\cal T} \equiv
\langle V_1(k_1) V_2(k_2) V_3(k_3) V_4(k_4)\rangle
= 0,
\end{equation}
where $V_1$ is the vertex operator
of a zero-norm state,
while $V_2, V_3$ and $V_4$ are vertex operators
of 3 arbitrary physical states.
Taking $V_1$ to be at this mass level,
and is thus of the form (\ref{mass2state}),
we can decompose ${\cal T}$ as
\begin{equation}
{\cal T} = \ep_{\mu\nu} {\cal T}^{\mu\nu}
+\ep_{\mu}{\cal T}^{\mu},
\end{equation}
where ${\cal T}^{\mu\nu}$ and ${\cal T}^{\mu}$
are the correlation functions for $V_1$
being the basis states
$\alpha_{-1}^{\mu}\alpha_{-1}^{\nu}|0;k\rangle$
and $\alpha_{-2}^{\mu}|0;k\rangle$
\begin{equation}
{\cal T}^{\mu\nu} \leftrightarrow
V_1 = \del X^{\mu}\del X^{\nu} e^{ik_1\cdot X},
\qquad
{\cal T}^{\mu}\leftrightarrow
V_1 = \del^2 X^{\mu} e^{ik_1\cdot X}.
\end{equation}
Strictly speaking, ${\cal T}^{\mu\nu}$
and ${\cal T}^{\mu}$ are not well-defined
because only physical states admit a path integral
independent of the gauge-fixing condition.
Hence we should restrict our attention
to those linear combinations
that correspond to physical states.

The decoupling of the zero-norm states
(\ref{ZNS11}) and (\ref{ZNS12}) from
other physical states implies that
\begin{eqnarray}
&\sqrt{2} {\cal T}^{LP}+{\cal T}^L=0,
\label{WI11} \\
&\sqrt{2} {\cal T}^{TP}+{\cal T}^T=0, \\
&{\cal T}^\m{}_\m+6{\cal T}^{PP}
+5\sqrt{2}{\cal T}^P=0, \label{WI13}
\end{eqnarray}
where $P, L, T$ stand for contraction with
the polarization vectors $e^P, e^L, e^T$,
respectively.
We will refer to these relations
as Ward identities.

For higher and higher mass levels,
zero-norm states are more and more complicated,
and so are the corresponding Ward identities.
But since we will only focus on the high energy limit,
there is an easier way to derive the Ward identities.
When we take the high energy limit of a zero-norm state,
it will not be of zero norm anymore
because we will ignore higher order components
in the $1/E$ expansion. 
Thus we should simply consider
the decoupling of the spurious states.
The Ward identities derived from spurious states are
\begin{eqnarray}
&\sqrt{2} {\cal T}^{\m P}+{\cal T}^\m=0,
\label{SS11} \\
&\frac{1}{2}{\cal T}^\m{}_\m+\sqrt{2}{\cal T}^P=0.
\label{SS12}
\end{eqnarray}
Notice that eqs.(\ref{WI11}-\ref{WI13})
are linear combinations of eqs.(\ref{SS11})
and (\ref{SS12}).
%% new 
The spurious states %constitute
lead to a larger %symmetry
set of constraints.
For instance they lead to a constraint
($\sqrt{2} {\cal T}^{PP}+{\cal T}^P = 0$)
which does not exist in eqs.(\ref{WI11}-\ref{WI13}).

Now we assume that scattering amplitudes remain
the same at the leading order in the high energy limit
under the replacement
\begin{equation}
\label{replace}
e^P \rightarrow e^L.
\end{equation}
Naively, this seems a direct result
of the fact that $e^P - e^L = {\cal O}(1/E)$.
However, if the scattering amplitude
does not have a smooth high energy limit,
this assumption may not be correct.
The validity of the replacement (\ref{replace})
is an assertion of the smoothness of
the high energy limit of string theory \cite{CHL}.

Under the replacement (\ref{replace}),
eqs.(\ref{SS11}) and (\ref{SS12}) become
\begin{eqnarray}
&\sqrt{2} {\cal T}^{LP}+{\cal T}^L=0,
\quad \rightarrow \quad
\sqrt{2}{\cal T}^{LL} + {\cal T}^L = 0,
\label{LL}
\\ 
&\sqrt{2} {\cal T}^{TP}+{\cal T}^T=0,
\quad \rightarrow \quad
\sqrt{2}{\cal T}^{TL} + {\cal T}^T = 0,
\label{TL}
\\
&\frac{1}{2}(-{\cal T}^{PP}
+{\cal T}^{LL}+{\cal T}^{TT})
+\sqrt{2}{\cal T}^{P}=0
\quad \rightarrow \quad
\frac{1}{2}{\cal T}^{TT}+\sqrt{2}{\cal T}^L=0.
\label{TT}
\end{eqnarray}
Here we used the fact that
transverse polarizations $e^I$
are kinematically suppressed to rewrite
${\cal T}^{\mu}{}_{\mu}$ as
$-{\cal T}^{PP}+{\cal T}^{LL}+{\cal T}^{TT}$.
(After replacing $P$ by $L$,
${\cal T}^{\mu}{}_{\mu}$ becomes just ${\cal T}^{TT}$.)

For $M^2 = 2$,
the physical amplitudes at the leading order
are ${\cal T}^{LL}$ and ${\cal T}^{TT}$.
Note that, although $\a^L_{-1}\a^L_{-1}|0;k\rangle$
and $\a^T_{-1}\a^T_{-1}|0;k\rangle$ are not physical states,
they can be extended into physical states
without changing the 4-point function
\begin{eqnarray}
(\a^L_{-1}\a^L_{-1}-\a^I_{-1}\a^I_{-1})|0;k\rangle
\quad \rightarrow \quad {\cal T}^{LL}, \\
(\a^T_{-1}\a^T_{-1}-\a^I_{-1}\a^I_{-1})|0;k\rangle
\quad \rightarrow \quad {\cal T}^{TT}.
\end{eqnarray}

From eqs.(\ref{LL}) and (\ref{TT}),
we expect ${\cal T}^{LL}$ and ${\cal T}^{TT}$
to have the ratio $1:4$
in the high energy limit.
By directly calculating the exact amplitudes and
expanding them in powers of $1/E$,
we can verify this result.
As an example, the exact 4-point function
${\cal T}^{\mu\nu}$ with $V_2, V_3, V_4$
corresponding to 3 tachyons is
\begin {eqnarray}
{\cal T}^{\m\n}&=&
\int_\infty^\infty \prod_{i=1}^4 dx_i
\langle :\del X^\m \del X^\n e^{ik_1\cdot X}::e^{ik_2 \cdot X}:
:e^{ik_3 \cdot X}::e^{ik_4 \cdot X}: \rangle \\
&=&
\left(
\frac{\Gamma(\frac{-s}{2}-1)\Gamma(\frac{-t}{2}-1)}{\Gamma(\frac{u}{2}+2)}
+\frac{\Gamma(\frac{-t}{2}-1)\Gamma(\frac{-u}{2}-1)}{\Gamma(\frac{s}{2}+2)}
+\frac{\Gamma(\frac{-u}{2}-1)\Gamma(\frac{-s}{2}-1)}{\Gamma(\frac{t}{2}+2)}
\right)
\nonumber \\
&&\times \left[
s/2(s/2+1)k^\m_3k^\n_3-2(s/2+1)(u/2+1)k^{(\m}_2k^{\n)}_3
+u/2(u/2+1)k^\m_2k^\n_2\right]+(k_3 \leftrightarrow k_4),
\label{Tmunu}
\end{eqnarray}
where $s=-(k_2+k_1)^2$, $t=-(k_2+k_3)^2$
and $u=-(k_2+k_4)^2$.

We can now calculate all amplitudes
in eq.(\ref{Tmunu}).
After some algebra, we get
\begin{eqnarray}
{\cal T}^{LL}&=&{\cal T}(2)E^6
[2\sin^2{\phi}+O(\frac{1}{E^4})],   \\
{\cal T}^{TT}&=&{\cal T}(2)E^6
[8\sin^2{\phi}+20\sin^2{\phi}\frac{1}{E^2}
                 +O(\frac{1}{E^4})],   \\
{\cal T}^{LT}&=&{\cal T}(2)E^5
[4\sqrt{2}\cos{\phi}\sin{\phi}+
      6\sqrt{2}\cos{\phi}\sin{\phi}\frac{1}{E^2}+O(\frac{1}{E^4})],
\end{eqnarray}
where
\begin{equation}
{\cal T}(2)=
\frac{\Gamma(\frac{-s}{2}-1)\Gamma(\frac{-t}{2}-1)}
{\Gamma(\frac{u}{2}+2)}
+\frac{\Gamma(\frac{-t}{2}-1)\Gamma(\frac{-u}{2}-1)}
{\Gamma(\frac{s}{2}+2)}       +\frac{\Gamma(\frac{-u}{2}-1)\Gamma(\frac{-s}{2}-1)}
{\Gamma(\frac{t}{2}+2)}.
\label{T2}
\end{equation}
We find that indeed
${\cal T}^{LL}:{\cal T}^{TT}=1:4$
in the high energy limit.

At this mass level,
${\cal T}^{LL}$ and ${\cal T}^{TT}$
are the only physical amplitudes at the leading order.
All other physical states are either
related to them via zero-norm states,
or are at a lower order.

It is interesting that eqs.(\ref{LL}) and (\ref{TT})
also determine ${\cal T}^L$
to have a fixed ratio with ${\cal T}^{LL}$
and ${\cal T}^{TT}$
\begin{equation}
{\cal T}^{LL}:{\cal T}^{TT}:{\cal T}^L
= 1 : 4 : -\sqrt{2},
\end{equation}
although
${\cal T}^L$ does not correspond to a physical state.
Under conformal transformations,
the path integral is not invariant
for non-physical states,
so it seems weird to predict a fixed relation
involving ${\cal T}^L$.
The reason is that the conformal anomaly
of ${\cal T}^L$ appears only at the subleading
order ($\frac{1}{E^2}$).
On the other hand,
for the amplitude ${\cal T}^T$
which is at a lower order than ${\cal T}^L$,
anomaly occurs at its leading order,
so it can not have a well defined value.

At this mass level,
there is only 1 amplitude ${\cal T}^{LT}$
at the next-to-leading order,
so we need to look at higher mass levels.

\section{Second massive level ($M^2 = 4$)}

The details of computation for the second massive level
($M^2 = 4$) is similar to that for the first massive level
in the previous section.
We simply list the results here.

The physical states of interest are
\begin{eqnarray}
|A\rangle &=& \left[
(\a^T_{-1})^3 - 3 (\a^I_{-1})^2\a^T_{-1}
\right]|0;k\rangle, \\
|B\rangle &=& \left[
\a^T_{-1}(\a^L_{-1})^2 - \a^T_{-1}(\a^I_{-1})^2
\right]|0;k\rangle, \\
|C\rangle &=& \left[
\a^L_{-1}(\a^T_{-1})^2 - \a^L_{-1}(\a^I_{-1})^2
\right]|0;k\rangle, \\
|D\rangle &=& \left[
(\a^L_{-1})^3 - 3 \a^L_{-1}(\a^I_{-1})^2
\right]|0;k\rangle, \\
|E\rangle &=& \frac{1}{2} \left[
\a^T_{-1}\a^L_{-2} - \a^L_{-1}\a^T_{-2}
\right]|0;k\rangle.
\end{eqnarray}
The 2nd terms on the right hand side
can be ignored in the high energy limit.

With the following notation for
amplitudes associated with different choices of $V_1$
\begin{eqnarray}
{\cal T}^{\mu\nu\lam} &\leftrightarrow&
\a^{\mu}_{-1}\a^{\nu}_{-1}\a^{\lam}_{-1} |0;k\rangle, \\
{\cal T}^{\mu\nu} &\leftrightarrow&
\a^{\mu}_{-1}\a^{\nu}_{-2}|0;k\rangle, \\
{\cal T}^{\mu} &\leftrightarrow&
\a^{\mu}_{-3}|0;k\rangle,
\end{eqnarray}
the decoupling of spurious states implies,
after replacing $P$ by $L$,
\begin{eqnarray}
&{\cal T}^{LTT}+{\cal T}^{TT} = 0,
\label{a} \\
&{\cal T}^{LLT}+{\cal T}^{(LT)} = 0, \\
&{\cal T}^{LLL}+{\cal T}^{LL} = 0,
\label{c} \\
&{\cal T}^{LT}+{\cal T}^T = 0, \\
&{\cal T}^{LL}+{\cal T}^L = 0,
\label{e} \\
&{\cal T}^{TT}+2{\cal T}^L = 0, \\
&\frac{1}{2}{\cal T}^{TTT}+2{\cal T}^{TL}+{\cal T}^T = 0, \\
&\frac{1}{2}{\cal T}^{LTT}+2{\cal T}^{LL}+{\cal T}^L = 0,
\label{g}
\end{eqnarray}
where ${\cal T}^{(LT)} \equiv \frac{1}{2}
\left({\cal T}^{LT}+{\cal T}^{TL}\right)$.

Together with some rough power counting,
these linear relations allow us to determine
which amplitudes are at the leading order,
as well as their ratios.
It turns out that the states
$|A\rangle$, $|B\rangle$ and $|E\rangle$
are at the leading order and have the ratios
\cite{ChanLee1}
\begin{equation}
{\cal T}(A) : {\cal T}(B) : {\cal T}(E)
= 8: 1: -1.
\end{equation}
The ratios can be derived algebraically
from eqs.(\ref{a})-(\ref{g}).
Similar to the case of $M^2 = 2$,
there is another non-physical state
($\left[\alpha^T_{-1}\alpha^L_{-2}+
\alpha^L_{-1}\alpha^T_{-2}\right]|0;k\rangle$)
at $M^2 = 4$
which is linearly related to these states
at the leading order.

What is new is that the amplitudes
at the next-to-leading order are also identical
up to numerical constant factors
\begin{equation}
{\cal T}(C): {\cal T}(D) = 2: 1.
\end{equation}
This ratio can be solved from the Ward identities
(\ref{c}), (\ref{e}) and (\ref{g}) above.
Unlike the leading order case,
only physical amplitudes have fixed ratios with each other.
Because of this difference,
it is much harder to derive a master formula for
the ratio between any two next-to-leading order amplitudes
at arbitrary mass levels,
since we do not have general formulas for
physical states at arbitrary mass levels.

Utilizing the ratios among the leading order amplitudes,
one can construct physical states whose
4-point functions are next to the next-to-leading order
\begin{equation}
(|A\rangle - 8 |B\rangle), \qquad
(|E\rangle + 2 |B\rangle).
\end{equation}
Unfortunately, the ratio of their amplitudes
is not a constant.

\section{Third massive level ($M^2 = 6$)}

Analogous to the previous two cases,
we list all the physical states
that we should consider
\begin{eqnarray}
|A\rangle &=& \left[
(\a^T_{-1})^4 - 6(\a^T_{-1})^2(\a^I_{-1})^2
+(\a^I_{-1})^4 \right]|0;k\rangle, \\
|B\rangle &=& \left[
(\a^T_{-1})^3\a^L_{-1} - 3\a^T_{-1}\a^L_{-1}(\a^I_{-1})^2
\right]|0;k\rangle, \\
|C\rangle &=& \left[
(\a^T_{-1})^2(\a^L_{-1})^2 - (\a^I_{-1})^2(\a^L_{-1})^2
- (\a^I_{-1})^2(\a^T_{-1})^2 + \frac{1}{3}(\a^I_{-1})^4
\right]|0;k\rangle, \\
|D\rangle &=& \left[
(\a^L_{-1})^3\a^T_{-1} - 3\a^L_{-1}\a^T_{-1}(\a^I_{-1})^2
\right]|0;k\rangle, \\
|E\rangle &=& \left[
(\a^L_{-1})^4 - 6 (\a^L_{-1})^2(\a^I_{-1})^2 + (\a^I{-1})^4
\right]|0;k\rangle, \\
|F\rangle &=& \left[
\a^T_{-1}\a^T_{-3} - \frac{3}{4}(\a^T_{-2})^2
- 3(\a^T_{-1})^2(\a^I_{-1})^2 + \frac{1}{2}(\a^I_{-1})^4
\right]|0;k\rangle, \\
|G\rangle &=& \left[
\a^T_{-1}\a^L_{-3}+\a^L_{-1}\a^T_{-3}
- \frac{3}{2}\a^T_{-2}\a^L_{-2}
- 6\a^T_{-1}\a^L_{-1}(a^I_{-1})^2
\right]|0;k\rangle, \\
|H\rangle &=& \left[
\a^L_{-1}\a^L_{-3} - \frac{3}{4}(\a^L_{-2})^2
- 3(\a^L_{-1})^2(\a^I_{-1})^2 + \frac{1}{2}(\a^I_{-1})^4
\right]|0;k\rangle, \\ 
|I\rangle &=& \frac{1}{2}\left[
(\a^T_{-1})^2\a^L_{-2}-(\a^I_{-1})^2\a^L_{-2}
-\a^T_{-1}\a^L_{-1}\a^T_{-2}+\a^I_{-1}\a^L_{-1}\a^I_{-2}\right]|0;k\rangle, \\
|J\rangle &=& \frac{1}{2}\left[
(\a^L_{-1})^2\a^T_{-2}-(\a^I_{-1})^2\a^T_{-2}
-\a^L_{-1}\a^T_{-1}\a^L_{-2}+\a^I_{-1}\a^T_{-1}\a^I_{-2}\right]|0;k\rangle .
\end{eqnarray}

To skip lengthy details,
let us just give the final result.
Solving the constraint obtained from decoupling
spurious states (and replacing $e^P$ by $e^L$),
one can find the ratios among amplitudes
at the leading order \cite{ChanLee1}
\begin{equation}
{\cal T}(A): {\cal T}(C): {\cal T}(E): {\cal T}(H): {\cal T}(I)
= 96: 8: 2: -3: -4\sqrt{6}.
\end{equation}
For example, 
\begin {equation}
{\cal T}(A)={\cal T}(6)E^{12}\left[32\sin^4\phi-160\sin^4\phi\frac{1}{E^2}+O(\frac{1}{E^4})\right],
\end {equation}
where
\begin{equation}
{\cal T}(6)=
\frac{\Gamma(\frac{-s}{2}-1)\Gamma(\frac{-t}{2}-1)}
{\Gamma(\frac{u}{2}+2)}
+\frac{\Gamma(\frac{-t}{2}-1)\Gamma(\frac{-u}{2}-1)}
{\Gamma(\frac{s}{2}+2)}       +\frac{\Gamma(\frac{-u}{2}-1)\Gamma(\frac{-s}{2}-1)}
{\Gamma(\frac{t}{2}+2)}.
\label{T6}
\end{equation}
(Notice that ${\cal T}(6)$ has the same form as ${\cal T}(2)$ in eq.(\ref{T2}),
but the definitions of s, t, and u depend on the mass levels.) 
Again, there is another non-physical state such as 
$(\a^T_{-1})^2\a^L_{-2}|0;k\rangle$ which is linearly related to these 
states at the leading order. \\
A new result of this paper is the ratio among
the next-to-leading order amplitudes
\begin{equation}
{\cal T}(B): {\cal T}(D): {\cal T}(G): {\cal T}(J)
= 36: 11: 3: 2\sqrt{6}.
\end{equation}
For the 4-point functions involving 3 tachyons,
we have
\begin{equation}
{\cal T}(B)={\cal T}(6)E^{11}\left[16\sqrt{6}\cos\phi\sin^3\phi
-\frac{64\sqrt{6}\cos\phi\sin^3\phi}{E^2}+O(\frac{1}{E^4})\right],
\end{equation}
etc., and the ratios above are verified.

Unlike the leading order amplitudes,
there is no non-physical state whose amplitude
has a definite ratio with any physical amplitude.

The next-to-leading order amplitudes are of order
$1/E$ smaller than the leading order amplitudes.
There are also amplitudes of order $1/E^2$ smaller
than the leading order amplitudes, such as
those for $|F\rangle$, $(|A\rangle - 12|C\rangle)$, $\sqrt{6}|I\rangle+3|C\rangle$
and $(|C\rangle-4|E\rangle)$.
They are not all proportional to each other, but they are not independent either:
\begin{eqnarray}
{\cal T}(F)&=&{\cal T}(6)E^{10}\left[8\sin^2\phi-\frac{24\sin^2\phi}{E^2}
               +O(\frac{1}{E^4})\right], \label{F} \\
{\cal T}(A-12C)&=&{\cal T}(6)E^{10}\left[-16(33\cos^2\phi+1)\sin^2\phi
      +\frac{4(193+191\cos(2\phi)\sin^2\phi}{E^2}+O(\frac{1}{E^4})\right], \\
{\cal T}(C-4E)&=&{\cal T}(6)E^{10}\left[\frac{-4(73\cos^2\phi-1)\sin^2\phi}{3}
      +\frac{2(151+169\cos(2\phi)\sin^2\phi)}{3E^2}+O(\frac{1}{E^4})\right], \\
{\cal T}(\sqrt{6}I+3C)&=&{\cal T}(6)E^{10}\left[144\cos^2\phi\sin^2\phi
      -\frac{6(35+37\cos(2\phi))\sin^2\phi}{E^2}+O(\frac{1}{E^4})\right].
      \label{IC}
\end{eqnarray}
We can see that in the 1/E expansion, 
there are only 2 degrees of freedom at the leading order which are proportional to 
$\cos^2\phi\sin^2\phi$ and $\sin^2\phi$.
This result is consistent with the DDF gauge consideration (see below):
there are only 2 independent states 
$A^T_{-1}A^T_{-3}|0;k\rangle$ and $A^T_{-2}A^T_{-2}|0;k\rangle$
at the 2nd subleading order.

\section{Comments on the linear relations}
\label{comment}

The fact that we can relate all amplitudes
at the leading order to each other
(for a given mass level) implies that
it is possible to choose another basis of
physical states such that there is
a unique state in the basis at the leading order,
with all other states in the basis
being subleading.
This basis was found in \cite{CHLTY3}
and called {\em DDF gauge}.
A generic state in the basis looks like
\begin{equation}
\label{DDFstate}
A_{-n_1}^{i_1}\cdots A_{-n_m}^{i_m}|p_0\rangle,
\end{equation}
where the $A_{-n}^i$'s denote DDF operators. 
To define the DDF operators,
one has to choose a light-like vector $k$.
Here we choose $k$ to be proportional to
$(e^L - e^P)$.
For the state to describe a particle
with momentum $p$,
we should choose the parameter $p_0$
in (\ref{DDFstate}) to be
$p_0 = p - (\sum_i n_i) k$,
which is a linear combination of $e^L$ and $e^P$.
The polarization $e^i$ of a DDF operator $A^i_{-n}$
can only be one of the spatial directions
transverse to the momentum.

An important property of the states
in the DDF gauge is that it is only composed
of the creation operators $\a^i_{-n}$
and $\a^{L-P}_{-n}$ when we expand them
in terms of the usual creation operators $\a^{\mu}_{-n}$.
In the computation of correlation functions,
we contract all Lorentz indices.
Because the components in the polarization vector
$e^{L-P} \equiv e^L - e^P$ is of order ${\cal O}(E^{-1})$,
while those of $e^T$ is of order ${\cal O}(E^0)$,
each factor of $\a^{L-P}_{-n}$ contributes
a relatively suppressed amplitude in
the high energy limit.
The conclusion of this consideration is
that a state created by $m_1$ DDF operators
dominates over another state created by
$m_2$ DDF operators if and only if $m_1 > m_2$.
The difference in their scaling behavior
in the high energy limit is simply $E^{m_1-m_2}$.
Therefore, for our computation of 4-point functions,
the unique state at the leading order
at mass level $n$ is
\begin{equation}
(A_{-1}^T)^n |0, p_0\rangle.
\end{equation}
Correspondingly,
there is a unique 4-point function
at the leading order for each mass level
(up to an overall constant factor
given by (\ref{TheRatio}).

The state at the next-to-leading order
is also unique
\begin{equation}
(A_{-1}^T)^{n-2}A_{-2}^T |0, p_0\rangle.
\end{equation}
This implies that all 4 point functions
at the next-to-leading order
must also be proportional to each other.
This is what we showed explicitly for $M^2 = 4$ and $M^2 = 6$.
Less obvious is that all the ratios
among the 4-point functions can be
algebraically derived by setting
$e^P \rightarrow e^L$ in the spurious states.

At the 2nd subleading order
(next to the next-to-leading order),
there are two independent DDF states
\begin{equation}
(A_{-1}^T)^{n-3}A_{-3}^T |0, p_0\rangle 
\qquad \mbox{and} \qquad
(A_{-1}^T)^{n-4}(A_{-2}^T)^2 |0, p_0\rangle .
\end{equation}
Unless the two states happen to have
the same high energy behavior,
we do not expect all amplitudes at the 2nd subleading order
to be proportional to each other. 
They should satisfy linear relations
involving 3 states at a time,
as each state at this order is a linear combination
of the two states above plus
some other states at even lower orders.
This is indeed what we observed in eqs.(\ref{F})-(\ref{IC})
for the mass level $M^2 = 6$.
Similarly,
there can be nontrivial linear relations
among amplitudes at any order for
sufficiently high mass levels.
For example, at the next order
(the 3rd subleading order), we have 3 independent
amplitudes associated with the following DDF states:
\begin{eqnarray}
(A_{-1}^T)^{n-4}A_{-4}^T |0, p_0\rangle, \qquad
(A_{-1}^T)^{n-5}A_{-2}^T A_{-3}^T |0, p_0\rangle, \qquad
(A_{-1}^T)^{n-6}(A_{-2}^T)^3 |0, p_0\rangle.
\end{eqnarray}

After we learned that these linear relations
are in some sense ``trivialized'' in the DDF gauge,
a legitimate question is whether
all these infinitely many linear relations
among amplitudes have anything to do with
any symmetry at all.
The answer to this question is not obvious,
but let us try to give some hints.
Firstly, although the linear relations seem trivial
in the DDF gauge,
the existence of the DDF gauge is highly nontrivial
(for example, it exists only for $D=26$).
The nontrivial content of the linear relations
is part of what made the DDF gauge possible,
which tells us that string theory has the same
number of degrees of freedom as a theory of
massless higher spin gauge fields.

Furthermore, there is an ``empirical evidence''
supporting a close connection between
the linear relations and symmetry.
The evidence is the 2D string theory,
which has a $w_{\infty}$ symmetry
dictating all scattering amplitudes \cite{Winfinity}.
The generators of the $w_{\infty}$ symmetry
are vertex operators of the so-called discrete states.
It was shown \cite{CL} that the same $w_{\infty}$ algebra
is generated by a class of zero-norm states.
Furthermore, it was also
shown \cite{CHLTY1} that,
in the high energy limit,
this class of zero-norm states approaches to
the discrete states
(also via the replacement $e^P\rightarrow e^L$),
and thus the conservation laws for each discrete state
is matched with the Ward identity of a zero-norm state.
This observation suggests that the linear relations
derived from the decoupling of zero-norm states
could bear some resemblance with
the algebraic structure of the hidden symmetry.

\section{Generic mass levels}

In this section we will give an explicit expression for
the next-to-leading order amplitudes for all mass levels
up to an overall constant,
and will establish a remarkably simple connection
between the mass levels
and the functional dependence on
the scattering angle $\phi$.
Since all these amplitudes at the same mass level
are proportional to each other,
we only need to consider a representative physical state
at the next-to-leading order for each mass level.

The representative physical state we choose
to compute is of the form
\begin{equation}
\label{stateansatz}
\left[
\alpha_{-1}^L(\alpha_{-1}^T)^{n-1}+
\sum_{m=1}^{n-1}a_m\alpha_{-1}^L(\alpha_{-1}^T)^{n-1-m}
(\alpha_{-1}^{I})^m \right]|0;k\rangle,
\end{equation}
where the numerical parameters
$a_k$ should be chosen such that
this is a physical state.
Since the state only involves creation operators
at level 1 ($\alpha_{-1}^{\mu}$),
almost all Virasoro generators $L_n$ with $n>0$
trivially annihilate the state.
The only exception is the term
$\frac{1}{2}\alpha_{1}\cdot\alpha_{1}$ in $L_2$.
Thus it is only the traceless condition that
needs to be taken care of.
One can easily convince oneself that
there are sufficient parameters $a_k$'s
for this purpose.
For our computation of 4-point functions
in the high energy limit,
the values of $a_k$'s are irrelevant.
We focus our attention on the first term
in (\ref{stateansatz}).

The tree level 4-point function is
an integral of
\begin{equation}
{\cal A} = |y_{12}||y_{13}||y_{23}|
\langle V_1(y_1) V_2(y_2) V_3(y_3) V_4(y_4) \rangle_{D_2}
\end{equation}
over the moduli space.
The factor $|y_{12}||y_{13}||y_{23}|$
is given by the ghost part of the vertices,
and $V_i$'s stand for the matter part.
The vertex operators we will compute are of the form
\begin{eqnarray}
V_i(y_i) = \left[\prod_{a=1}^{n_i} \del X^{\mu_a}(y_i)\right]
e^{ik_i\cdot X(y_i)},
\qquad i = 1, 2, 3, 4,
\end{eqnarray}
where normal ordering is to be carried out.

There is a useful formula \cite{Polchinski}
to evaluate this correlation function at tree level
\begin{equation}
\langle \prod_{i=1}^n e^{ik_i\cdot X(y_i)}
\prod_{a=1}^p \del_y X^{\mu_a}(y'_a)\rangle_{D_2}
=iC (2\pi)^{26}\delta^{(26)}(\sum_i k_i)
\prod_{i<j}|y_{ij}|^{k_i\cdot k_j}
\langle\prod_a\left[v^{\mu_a}(y'_a)
+q^{\mu_a}(y'_a)\right]\rangle_{D_2},
\label{Pol}
\end{equation}
where
\begin{equation}
v^{\mu}(y) = -i\sum_{i}\frac{k_i^{\mu}}{y-y_i}
\end{equation}
and the $q$'s are contracted using
\begin{equation}
\langle q^{\mu}(y)q^{\nu}(y')\rangle =
-\frac{\eta^{\mu\nu}}{(y-y')^2}.
\end{equation}

The calculation of the 4-point functions
is straightforward but tedious.
Here we briefly describe the techniques we use.
We take the usual gauge-fixing condition
\begin{equation}
y_1=0, \quad y_2=1, \quad y_3\rightarrow \infty,
\quad y_4=x,
\end{equation}
where $x$ is the only modular parameter
to be integrated over.
A 4-point function is always a linear combination
of integrals of the form
\begin{equation}
\int_{-\infty}^{\infty}dx x^A (1-x)^B,
\end{equation}
where $A$ and $B$ are some of the Mandelstam variables
$s, t, u$ plus constants.
This integral can be decomposed into integrals over
3 different regions $(-\infty, 0), (0, 1)$
and $(1, \infty)$.
We have
\begin{equation}
\int_0^1 dx x^A (1-x)^B =
\frac{\Gamma(A+1)\Gamma(B+1)}{\Gamma(A+B+2)},
\end{equation}
and a change of variable $x = 1-1/x'$ gives
\begin{equation}
\int_{-\infty}^0 dx x^A (1-x)^B = (-1)^A
\frac{\Gamma(-A-B-1)\Gamma(A+1)}{\Gamma(-B)},
\end{equation}
and yet another change of variable $x = 1/x'$ gives
\begin{equation}
\int_1^{\infty} dx x^A (1-x)^B = (-1)^B
\frac{\Gamma(-A-B-1)\Gamma(B+1)}{\Gamma(-A)}.
\end{equation}
Using the following identity for the Gamma function
\begin{equation}
\Gamma(A+1) = A\Gamma(A),
\end{equation}
we find that, for all 3 integrals,
the effect of increasing the parameter $A$
to $A+1$ is equivalent to adding
a multiplicative factor of
\begin{equation}
\left(\frac{A+1}{A+B+2}\right).
\end{equation}
Thanks to this observation,
we can write the 4 point function
in the form
\begin{equation}
{\cal A} = iC (2\pi)^{26}\delta^{(26)}(\sum_i k_i)
\left[
\int_{-\infty}^{\infty} x^{-t/2-2}(1-x)^{-u/2-2}
\right]
{\cal B}(s,t,u),
\label{AB}
\end{equation}
where ${\cal B}(s,t,u)$ is a fraction of
polynomials of $s, t, u$.
The Mandelstam variables are linearly dependent:
\begin{equation}
s + t + u = \sum_{i=1}^4 m_i^2 = 2 (n-4),
\end{equation}
where $m_i=\sqrt{2(n_i-1)}$ is the mass
of the $i$-th particle,
and $n = \sum_{i=1}^4 n_i$ is the sum
over mass levels.

The common factor $\left[\cdot\right]$
in eq.(\ref{AB}) can be
written down more explicitly as
\begin{eqnarray}
\left[
\int_{-\infty}^{\infty} x^{-t/2-2}(1-x)^{-u/2-2}
\right] &=&
\frac{\Gamma(-\frac{t}{2}-1)\Gamma(-\frac{u}{2}-1)}
{\Gamma(\frac{s}{2}-n+2)} +
\nonumber \\
&&
\frac{\Gamma(-\frac{s}{2}+n-1)\Gamma(-\frac{t}{2}-1)}
{\Gamma(\frac{u}{2}+2)} +
\frac{\Gamma(-\frac{s}{2}+n-1)\Gamma(-\frac{u}{2}-1)}
{\Gamma(\frac{t}{2}+2)}.
\label{GGG}
\end{eqnarray}
As a digression we comment that the Sterling's formula
\begin{equation}
\Gamma(n+1)\simeq \sqrt{2\pi} n^{n+1/2} e^{-n}
\end{equation}
is valid only for $n>0$.
The Gamma function diverges at negative integers.
The expression
\begin{equation}
\Gamma(-x) = \frac{\pi}{-x\sin(\pi x)\Gamma(x)}
\end{equation}
allows us to write down a valid approximation
for large negative values.
However, since this common factor is common to
{\em all} amplitudes,
this complication is unnecessary
for our purpose of examining the relations among
scattering amplitudes.
We only need the high energy expansion of ${\cal B}$.

Let us first recall that when all vertices
are at the leading order, e.g.,
\begin{equation}
\label{stateleading}
V_i = (\del X^T)^{n_i} e^{ik_i\cdot X},
\end{equation}
the 4 point function has a simple high energy limit with
\begin{equation}
{\cal B} \simeq (-1)^{n_1+n_2}
(-E\sin\phi)^n.
\label{leading}
\end{equation}

After lengthy calculation,
we find that,
if the first $k$ vertices ($0 \leq k \leq 4$)
correspond to states at the next-to-leading order,
e.g. (\ref{stateansatz}),
while the rest of the 4 vertices are
at the leading order, e.g. (\ref{stateleading}),
the high energy limit of ${\cal B}$ is
\begin{equation}
{\cal B} \simeq (-1)^{n_1+n_2}
(-E\sin\phi)^{n-k} \prod_{a=1}^k
\left(-\frac{m_a}{2}\cos\phi\right).
\label{nexttoleading}
\end{equation}

We can summarize the expression above
by the following rule
\begin{eqnarray}
\del X^T \rightarrow -E\sin\phi,
\label{assoc-T} \\
\del X^L \rightarrow -\frac{m}{2}\cos\phi.
\label{assoc-L}
\end{eqnarray}
That is, apart from the common factor
of the 4-point function,
for every factor of $\del X^T$,
regardless of which vertex it resides in,
we associate a factor of $(-E\sin\phi)$.
The association of $\del X^L$
with the factor of $(-\frac{m}{2}\cos\phi)$
needs further explanation.
We only considered the case when
a vertex involves at most a single factor of $\del X^L$.
Vertices with an even number of $\del X^L$
and an arbitrary number of $\del X^T$
is at the leading order,
and vertices with an odd number of $\del X^L$
and an arbitrary number of $\del X^T$
is at the next-to-leading order.
The association is restricted
to states either of the form (\ref{stateansatz})
or of the form (\ref{stateleading}).
Nevertheless, we can now write down
the general expression of
all next-to-leading order amplitudes
at any mass level
(up to numerical constant factors).

Our factorization rules (\ref{assoc-T})
and (\ref{assoc-L}) can also be written
in terms of the DDF operators as
\begin{equation}
\label{factorule}
A_{-1}^T \rightarrow -E\sin\phi, \qquad
A_{-2}^T \rightarrow -\frac{m}{2}\cos\phi.
\end{equation}

In our computation, we note that
when there are more vertices involving $\del X^L$,
there is more cancellation in
the $1/E$ expansion of the product $\prod(v+q)$
in (\ref{Pol})
since the naive power counting would give
$\del X^L \sim E^2$ (which is wrong).
As a result the computation is
more complicated because
we need to take into consideration
higher and higher order terms in the $1/E$ expansion.
The many cancellations not only reduce
the 4-point function to a lower order,
but also lead to a result which is
consistent with
the remarkably simple factorization rules
(\ref{assoc-T}) and (\ref{assoc-L}).
The high energy limit of string theory
demonstrates a much simpler structure
than the theory at finite energy.

It is also interesting to note that,
up to a sign,
both the leading and next-to-leading amplitudes
(see eqs.(\ref{AB}), (\ref{GGG}), (\ref{leading}
and (\ref{nexttoleading}))
depend only on the sum of mass levels $n$,
instead of depending on all 4 numbers
$(n_1, n_2, n_3, n_4)$
(assuming that we use $E$ and $\phi$
as the parameters).
This is a feature valid only
in the high energy limit.
Its physical meaning remains to be understood.

\section*{Appendix}

Here we list our definition of the kinematic variables
involved in a 4-point function.
In Fig. 1,
we take the scattering plane
to be the $X^{1}-X^{2}$ plane.
The momenta of the particles are
\begin{align}
k_{1}  &  = (\sqrt{p^{2} + m_{1}^{2}}, -p, 0),\\
k_{2}  &  = (\sqrt{p^{2} + m_{2}^{2}}, p, 0),\\
k_{3}  &  = (-\sqrt{q^{2} + m_{3}^{2}}, -q\cos\phi, -q\sin\phi),\\
k_{4}  &  = (-\sqrt{q^{2} + m_{4}^{2}}, q\cos\phi, q\sin\phi).
\end{align}
They satisfy $k_{i}^{2} = -m_{i}^{2}$. In the high-energy limit, the
Mandelstam variables are
\begin{align}
s  &  \equiv-(k_{1} + k_{2})^{2} \equiv 4 E^{2},\\
t  &  \equiv-(k_{2} + k_{3})^{2},\\
u  &  \equiv-(k_{1} + k_{3})^{2}.
\end{align}
The polarization vectors for the 4 particles are
\begin{eqnarray}
&e^{L}(1) = \frac{1}{m_{1}} (p, - \sqrt{p^{2} + m_{1}^{2}}, 0),
\qquad  e^{T}(1) = (0, 0, - 1),\\
&e^{L}(2) = \frac{1}{m_{2}} (p, \sqrt{p^{2} + m_{2}^{2}}, 0),
\qquad  e^{T}(2) = (0, 0, 1),\\
&e^{L}(3) = \frac{1}{m_{3}}
(-q, - \sqrt{q^{2} + m_{3}^{2}} \cos\phi,
 -\sqrt{q^{2} + m_{3}^{2}} \sin\phi),
\qquad  e^{T}(3) = (0, - \sin\phi, \cos\phi),\\
&e^{L}(4) = \frac{1}{m_{4}} (-q, \sqrt{q^{2} + m_{4}}\cos\phi,
 \sqrt{q^{2} + m_{4}^{2}} \sin\phi),
\qquad  e^{T}(4) = (0, \sin\phi, -\cos\phi).
\end{eqnarray}

\begin{figure}[ptb]
\setlength{\unitlength}{2pt}
\par
\begin{center}
\begin{picture}(100,100)(-50,-50)
{\large
% draw k_i
\put(45,0){\vector(-1,0){42}} \put(-45,0){\vector(1,0){42}}
\put(2,2){\vector(1,1){30}} \put(-2,-2){\vector(-1,-1){30}}
\put(25,2){$k_1$} \put(-27,2){$k_2$} \put(11,20){$-k_3$}
\put(-24,-15){$-k_4$}
% draw e^T^i
\put(40,0){\vector(0,-1){10}} \put(-40,0){\vector(0,1){10}}
\put(26,26){\vector(-1,1){7}} \put(-26,-26){\vector(1,-1){7}}
\put(36,-16){$e^{T}(1)$} \put(-44,15){$e^{T}(2)$}
\put(15,36){$e^{T}(3)$} \put(-18,-35){$e^{T}(4)$}
% draw \phi_{CM}
\qbezier(10,0)(10,4)(6,6) \put(12,4){$\phi$}
% caption
\put(-43,-45){Fig.1 Kinematic variables
in the center of mass frame} }
\end{picture}
\end{center}
\end{figure}

\section*{Acknowledgment}

The authors thank Chuan-Tsung Chan, Chong-Sun Chu,
Tohru Eguchi, Clifford Johnson, Hsien-chung Kao,
Jen-Chi Lee, Yutaka Matsuo, Nicolas Moeller,
Shunsuke Teraguchi, Peter West and Yi Yang
for helpful discussions.
This work is supported in part by
the National Science Council,
and the National Center for Theoretical Sciences
(NSC 94-2119-M-002-001), Taiwan, R.O.C.
and the Center for Theoretical Physics
at National Taiwan University.


\vskip .8cm
\baselineskip 22pt
\begin{thebibliography}{99}
\itemsep 0pt

\bibitem{GrossMende}
  D.~J.~Gross and P.~F.~Mende,
  ``String Theory Beyond The Planck Scale,''
  Nucl.\ Phys.\ B {\bf 303}, 407 (1988).
  %%CITATION = NUPHA,B303,407;%%
  D.~J.~Gross and P.~F.~Mende,
  ``The High-Energy Behavior Of String Scattering Amplitudes,''
  Phys.\ Lett.\ B {\bf 197}, 129 (1987).
  %%CITATION = PHLTA,B197,129;%%

\bibitem{Gross}
  D.~J.~Gross,
  ``High-Energy Symmetries Of String Theory,''
  Phys.\ Rev.\ Lett.\  {\bf 60}, 1229 (1988).
  %%CITATION = PRLTA,60,1229;%%

\bibitem{GrossManes}
  D.~J.~Gross and J.~L.~Manes,
  ``The High-Energy Behavior Of Open String Scattering,''
  Nucl.\ Phys.\ B {\bf 326}, 73 (1989).
  %%CITATION = NUPHA,B326,73;%%

\bibitem{ChanLee1}
  C.~T.~Chan and J.~C.~Lee,
  ``Stringy symmetries and their high-energy limits,''
  Phys.\ Lett.\ B {\bf 611}, 193 (2005)
  [arXiv:hep-th/0312226].
  %%CITATION = HEP-TH 0312226;%%

\bibitem{ChanLee2}
  C.~T.~Chan and J.~C.~Lee,
  ``Zero-norm states and high-energy symmetries
  of string theory,''
  Nucl.\ Phys.\ B {\bf 690}, 3 (2004)
  [arXiv:hep-th/0401133].
  %%CITATION = HEP-TH 0401133;%%

\bibitem{CHL}
C.~T.~Chan, P.~M.~Ho and J.~C.~Lee,
``Ward identities and
high-energy scattering amplitudes in string theory,''
Nucl.\ Phys.\ B
\textbf{708}, 99 (2005)
[arXiv:hep-th/0410194].
%%CITATION = HEP-TH 0410194;%%

\bibitem{Lee}
  J.~C.~Lee,
  ``New Symmetries Of Higher Spin States In String Theory,''
  Phys.\ Lett.\ B {\bf 241}, 336 (1990).
  %%CITATION = PHLTA,B241,336;%%
   J.~C.~Lee, B. A. Ovrut,
  ``Zero Norm States And Enlarged Gauge Symmetries Of Closed Bosonic String In
  Background Massive Fields,''
  Nucl.\ Phys.\ B {\bf 336}, 222 (1990).
  %%CITATION = NUPHA,B336,222;%%
  J.~C.~Lee,
  ``Decoupling Of Degenerate Positive Norm States
  In String Theory,''
  Phys.\ Rev.\ Lett.\  {\bf 64}, 1636 (1990).
  %%CITATION = PRLTA,64,1636;%%
  J.~C.~Lee,
  ``Superconformal deformation, zero norm states
  and space-time symmetries of the superstring theories,''
  Z.\ Phys.\ C {\bf 54}, 283 (1992).
  %%CITATION = ZEPYA,C54,283;%%
  J.~C.~Lee,
  ``Heterotic massive Einstein-Yang-Mills-type symmetry
  and Ward identity,''
  Phys.\ Lett.\ B {\bf 337}, 69 (1994)
  [arXiv:hep-th/0503004].
  %%CITATION = HEP-TH 0503004;%%
  J.~C.~Lee,
  ``Spontaneously broken symmetry in string theory,''
  Phys.\ Lett.\ B {\bf 326}, 79 (1994)
  [arXiv:hep-th/0503056].
  %%CITATION = HEP-TH 0503056;%%
  J.~C.~Lee,
  ``Generalized On-Shell Ward Identities in String Theory,''
  Prog.\ Theor.\ Phys.\  {\bf 91}, 353 (1994)
  [arXiv:hep-th/0503005].
  %%CITATION = HEP-TH 0503005;%%
  H.~C.~Kao and J.~C.~Lee,
  ``Decoupling of degenerate positive-norm states
  in Witten's string field theory,''
  Phys.\ Rev.\ D {\bf 67}, 086003 (2003)
  [arXiv:hep-th/0212196].
  %%CITATION = HEP-TH 0212196;%%
  J.~C.~Lee,
  ``Zero-norm states and reduction of
  stringy scattering amplitudes,''
  Prog.\ Theor.\ Phys.\  {\bf 114}, 259 (2005)
  [arXiv:hep-th/0302123].
  %%CITATION = HEP-TH 0302123;%%

\bibitem{ChanLee3}
  C.~T.~Chan and J.~C.~Lee,
  ``One-loop massive scattering amplitudes
  and Ward identities in string theory,''
  Prog.\ Theor.\ Phys.\  {\bf 115}, 229 (2006)
  [arXiv:hep-th/0411212].
  %%CITATION = HEP-TH 0411212;%%

\bibitem{CHLTY1}
  C.~T.~Chan, P.~M.~Ho, J.~C.~Lee, S.~Teraguchi and Y.~Yang,
  ``Solving all 4-point correlation functions
  for bosonic open string theory in the high energy limit,''
  [arXiv:hep-th/0504138].
  %%CITATION = HEP-TH 0504138;%%

\bibitem{CHLTY2}
  C.~T.~Chan, P.~M.~Ho, J.~C.~Lee, S.~Teraguchi and Y.~Yang,
  ``High-energy zero-norm states and
  symmetries of string theory,''
  to appear in Phys. Rev. Lett.
  [arXiv:hep-th/0505035].
  %%CITATION = HEP-TH 0505035;%%

\bibitem{CHLTY3}
  C.~T.~Chan, P.~M.~Ho, J.~C.~Lee, S.~Teraguchi and Y.~Yang,
  ``Comments on the high energy limit of
  bosonic open string theory,''
  [arXiv:hep-th/0509009].
  %%CITATION = HEP-TH 0509009;%%

\bibitem{CLY}
  C.~T.~Chan, J.~C.~Lee and Y.~Yang,
  ``High energy scattering amplitudes of superstring theory,''
  Nucl.\ Phys.\ B {\bf 738}, 93 (2006)
  [arXiv:hep-th/0510247].
  %%CITATION = HEP-TH 0510247;%%

\bibitem{CHLTY4}
  C.~T.~Chan, P.~M.~Ho, J.~C.~Lee, S.~Teraguchi and Yi-Yang,
  ``Zero-norm states and stringy symmetries,''
  [arXiv:hep-th/0511283].
  %%CITATION = HEP-TH 0511283;%%

\bibitem{Moore}
  G.~W.~Moore,
  ``Finite in all directions,''
  [arXiv:hep-th/9305139].
  %%CITATION = HEP-TH 9305139;%%
  G.~W.~Moore,
  ``Symmetries and symmetry breaking in string theory,''
  arXiv:hep-th/9308052.
  %%CITATION = HEP-TH 9308052;%%

\bibitem{West}
  P.~C.~West,
  ``A Brief Review Of The Group Theoretic Approach
  To String Theory,''
  in ``Conformal Field Theory and Related Topics,''
  Proceedings of Third Annecy Meeting on Theoretical Physics,
  LAPP, Annecy le Vieux, France,
  Nucl. Phys. B (Proc. Suppl) {\bf 5B} (1988) 217.
  N.~Moeller and P.~West,
  ``Arbitrary four string scattering
  at high energy and fixed angle,''
  Nucl.\ Phys.\ B {\bf 729}, 1 (2005)
  [arXiv:hep-th/0507152].
  %%CITATION = HEP-TH 0507152;%%
  
\bibitem{WS}
See, for instance,
J.~Isberg, U.~Lindstrom, B.~Sundborg and G.~Theodoridis,
``Classical and quantized tensionless strings,''
Nucl.\ Phys.\ B \textbf{411}, 122 (1994)
[arXiv:hep-th/9307108];
%%CITATION = HEP-TH 9307108;%%
B.~Sundborg,
``Stringy gravity,
interacting tensionless strings and massless higher spins,''
Nucl.\ Phys.\ Proc.\ Suppl.\ \textbf{102}, 113 (2001)
[arXiv:hep-th/0103247];
%%CITATION = HEP-TH 0103247;%%
E.~Sezgin and P.~Sundell,
``Massless higher spins and holography,''
Nucl.\ Phys.\ B \textbf{644}, 303 (2002)
[Erratum-ibid.\ B \textbf{660}, 403 (2003)]
[arXiv:hep-th/0205131];
%%CITATION = HEP-TH 0205131;%%
C.~S.~Chu, P.~M.~Ho and F.~L.~Lin,
``Cubic string field theory in pp-wave background
and background independent Moyal structure,''
JHEP \textbf{0209}, 003 (2002)
[arXiv:hep-th/0205218].
%%CITATION = HEP-TH 0205218;%%

\bibitem{HS}
See, for instance,
C.~Fronsdal,
``Massless Fields With Integer
Spin,'' Phys.\ Rev.\ D \textbf{18}, 3624 (1978);
%%CITATION = PHRVA,D18,3624;%%
J.~Fang and C.~Fronsdal,
``Massless Fields With Half Integral Spin,''
Phys.\ Rev.\ D \textbf{18}, 3630 (1978);
%%CITATION = PHRVA,D18,3630;%%
M.~A.~Vasiliev,
``Progress in higher spin gauge theories,''
\textit{Prepared for 9th Marcel Grossmann Meeting
on Recent Developments in
Theoretical and Experimental General Relativity,
Gravitation and Relativistic Field Theories (MG 9),
Rome, Italy, 2-9 Jul 2000};
M.~A.~Vasiliev,
``Higher spin gauge theories in various dimensions,''
Fortsch.\ Phys.\ \textbf{52}, 702 (2004)
[arXiv:hep-th/0401177].
%%CITATION = HEP-TH 0401177;%%

\bibitem{Winfinity}
  J.~Avan and A.~Jevicki,
  ``Classical integrability and higher symmetries
  of collective string field theory,''
  Phys.\ Lett.\ B {\bf 266}, 35 (1991).
  %%CITATION = PHLTA,B266,35;%%
  I.~R.~Klebanov and A.~M.~Polyakov,
  ``Interaction of discrete states
  in two-dimensional string theory,''
  Mod.\ Phys.\ Lett.\ A {\bf 6}, 3273 (1991)
  [arXiv:hep-th/9109032].
  %%CITATION = HEP-TH 9109032;%%

\bibitem{CL}
  T.~D.~Chung and J.~C.~Lee,
  ``Discrete gauge states and W(infinity) charges in c = 1 2-d gravity,''
  Phys.\ Lett.\ B {\bf 350}, 22 (1995)
  [arXiv:hep-th/9412095].
  %%CITATION = HEP-TH 9412095;%%
  T.~D.~Chung and J.~C.~Lee,
  ``Superfield form of discrete gauge states in c = 1 2-d supergravity,''
  Z.\ Phys.\ C {\bf 75}, 555 (1997)
  [arXiv:hep-th/9505107].
  %%CITATION = HEP-TH 9505107;%%

\bibitem{Polchinski}
  J.~Polchinski,
  ``String theory. Vol. 1:
  An introduction to the bosonic string,''
  Cambridge University Press (1998).

\end{thebibliography}
\end{document}